\pdfoutput=1
\documentclass[conference]{IEEEtran}
\IEEEoverridecommandlockouts

\usepackage{amsmath,amssymb,amsfonts}
\usepackage{algorithmic}
\usepackage{graphicx}
\usepackage{textcomp}

\usepackage{enumitem}
\usepackage{verbatim}
\usepackage[nospace, nocompress]{cite}
\usepackage{multirow}
\usepackage{listings}
\usepackage{alltt}
\usepackage{fancyvrb}
\usepackage{soul}
\usepackage{authblk}
\makeatletter
\renewcommand\AB@affilsepx{, \protect\Affilfont}
\makeatother

\usepackage[singlelinecheck=false]{caption}

\newcommand{\boldhdr}[1]{\textbf{#1.}}

\newcommand\yida[1]{\textcolor{blue}{[yida: #1]}
}

\usepackage[table,xcdraw]{xcolor}
\usepackage{pifont}
\newcommand{\cmark}{\ding{51}}%
\usepackage{etoolbox}
\makeatletter
\patchcmd{\@makecaption}
  {\scshape}
  {}
  {}
  {}
\makeatother

\def\BibTeX{{\rm B\kern-.05em{\sc i\kern-.025em b}\kern-.08em
    T\kern-.1667em\lower.7ex\hbox{E}\kern-.125emX}}
\begin{document}

\title{Efficient Execution of Quantized Deep Learning Models: A Compiler Approach}

\author[1]{Animesh Jain}
\author[1]{Shoubhik Bhattacharya}
\author[2]{Masahiro Masuda}
\author[1]{Vin Sharma}
\author[1]{Yida Wang}
\affil[1]{Amazon Web Services}
\affil[2]{Edgecortix Inc.}

\maketitle

\begin{abstract}
A growing number of applications implement predictive functions using deep learning models, which require heavy use of compute and memory. For deep learning workloads to run well on a broad range of systems from cloud-scale clusters to low-power edge devices, they need to use available compute and memory resources more efficiently. One popular technique for increasing resource efficiency is 8-bit integer quantization, in which 32-bit floating point numbers (\texttt{fp32}) are represented using shorter 8-bit integer numbers. Although deep learning frameworks such as TensorFlow, TFLite, MXNet, and PyTorch enable developers to quantize models with only a small drop in accuracy, they are not well suited to execute quantized models on a variety of hardware platforms. For example, TFLite is optimized to run inference on ARM CPU edge devices but it does not have efficient support for Intel CPUs and Nvidia GPUs. In this paper, we address the challenges of executing quantized deep learning models on diverse hardware platforms by proposing an augmented compiler approach. A deep learning compiler such as Apache TVM can enable the efficient execution of model from various frameworks on various targets. Many deep learning compilers today, however, are designed primarily for \texttt{fp32} computation and cannot optimize a pre-quantized INT8 model. To address this issue, we created a new dialect called Quantized Neural Network (QNN) that extends the compiler's internal representation with a quantization context. With this quantization context, the compiler can generate efficient code for pre-quantized models on various hardware platforms. As implemented in Apache TVM, we observe that the QNN-augmented deep learning compiler achieves speedups of 2.35$\times$, 2.15$\times$, 1.35$\times$ and 1.40$\times$ on Intel Xeon Cascade Lake CPUs, Nvidia Tesla T4 GPUs, ARM Cortex-A CPUs on Raspberry Pi3 and Pi4 respectively against well optimized \texttt{fp32} execution. The use of QNN with compilation of pre-quantized models enables developers to achieve model execution performance comparable to the state-of-the-art framework-specific solutions but on a wider range of hardware platforms.
\end{abstract}

\section{Introduction}
\label{sec:intro}

\subsection{Background}
\label{sec:intro:back}
The effectiveness of deep learning in image processing and natural language processing tasks has led to the development of a growing number of applications that run deep learning models on a wide range of systems, from cloud-scale clusters to resource-limited edge-devices~\cite{bert, resnet, fb, tpu}. The widespread use of deep learning frameworks such as Tensorflow~\cite{tf}, PyTorch~\cite{pytorch} and MXNet~\cite{mxnet} drives the applications of deep learning. These frameworks enable model developers to quickly build, train, and deploy models on many hardware platforms. The frameworks provide a set of operators, where each operator represents a mathematical computation, e.g., convolution2D (referred to as conv2d), ReLU (rectified linear unit), batch normalization etc. These operators are typically converted to machine code using a hardware-specific library, e.g., Intel DNNL~\cite{mkl} and Nvidia CuDNN~\cite{cudnn}.

In general, deep learning models require substantial compute and memory resources~\cite{fb, tpu}, which can burden even powerful servers leave alone low-power edge devices. Researchers have implemented various techniques -- algorithms, software, hardware -- to reduce the compute and memory burden~\cite{mobilenet, tpu, tflite} of deep learning models to simplify their development and deployment~\cite{jiang2018efficient,wang2019unified}.

Among these techniques, quantization is a promising and well-studied approach. Quantization represents floating point 32-bit (\texttt{fp32}) numbers, which are used frequently in the deep learning models, with integer 8-bits (\texttt{int8})\cite{tflite, quant1, quant2}, reducing the memory footprint by a factor of four. The most widely-used form of integer quantization is uniform quantization~\cite{tflite}, where an \texttt{fp32} tensor ($A_{fp32}$) is represented with a quantized \texttt{int8} tensor ($Q_A$) along with quantization attributes - scale ($scale_A$) and zero point ($zp_A$) as shown below
\begin{equation}
A_{fp32} = scale_A * (Q_A - zp_A)
\label{eq:quant}
\end{equation}

\noindent Quantization enables a model to consume fewer compute and memory resources while keeping its accuracy close to that of the unquantized model (where the inputs and parameters are represented in \texttt{fp32}). We have observed that \texttt{int8} quantization is most widely used in the real world because (1) prior works have shown that \texttt{int8} representation works well empirically in preserving model accuracy~\cite{tflite, quant1, quant2}; (2) popular hardware platforms like Intel CPUs, Nvidia GPUs, and ARM CPUs are introducing low-level instructions support to perform \texttt{int8} data type computation efficiently.

As a result, deep learning frameworks have of late started to implement uniform \texttt{int8} quantization. Most of these efforts have focused on retaining accuracy~\cite{tflite, quant1, quant2}. For example, instead of scalar scale in Equation 1, we can use per-channel scale to get more fine-grained quantization~\cite{tflite_whitepaper}. Similarly, we can use symmetric or asymmetric quantization based on the value of zero points and choose a suitable performance-accuracy trade-off~\cite{tflite}.

However, there is very little focus on the broad deployment and efficient execution of these framework-quantized models (hereafter referred to as \emph{pre-quantized models}) on a variety of  platforms (server and edge). This paper tackles the challenges associated with model inference. Specifically, this paper presents a universally applicable approach to execute pre-quantized models from deep learning frameworks (e.g. TFLite, MXNet, PyTorch) efficiently on a variety of hardware platforms (e.g. Intel CPUs, Nvidia GPUs, ARM CPUs).

\subsection{Quantization in Deep Learning Frameworks}
\label{sec:intro:framework}
Adding support for the execution of pre-quantized models in each deep learning framework is not an efficient use of developer time. First, there are several popular frameworks in widespread use, which means that the same effort must be duplicated across multiple frameworks. Second, a model is trained and quantized in a framework, it can only run in the same framework on the hardware that the framework supports; i.e. an MXNet model quantized on Intel CPUs cannot run in TFLite on ARM CPUs. Third, a framework typically handles quantization by adding new quantized operators, e.g., TFLite has quantized operators such as quantized\_conv2d, quantized\_add. However, adding quantized operators to a framework does not automatically enable the framework to execute pre-quantized models efficiently. There are several obstacles facing a framework.
 
\begin{itemize}[leftmargin=*]

    \item \boldhdr{Lack of Kernel Library Support} Frameworks typically rely on high-performance kernel libraries (e.g. Intel DNNL and Nvidia CuDNN) to process computationally-intensive operators. When a new operator is added to the framework, the kernel libraries integrated with the framework must add corresponding support this new operator. Without that support, the quantized model either cannot run well or cannot run at all.
    
    \item \boldhdr{Per-operator Overhead} Operators in the framework are tightly coupled. The best use of a new operator requires corresponding rules and updates across the framework optimization toolchain. For example, one might want to fuse quantized conv2d with a requantize operator.
    
    \item \boldhdr{Diverse Hardware Platforms} Most critically, the hardware platforms have varying levels of support for quantization. Each platform often has specific requirements for these new operators, e.g., Intel CPUs with x86 architecture prefer the input data types of the quantized conv2d to be $uint8 \times int8$ due to the Intel Vector Neural Network Instructions (VNNI) requirement~\cite{vnni}. Similarly, CPUs in ARMv8 architecture have special instructions to accelerate the \texttt{int16} multiply-accumulate, while ARMv8.2 CPUs introduce DOT instruction to directly speed up \texttt{int8} multiply-accumulate~\cite{dot}. These requirements percolate up to the framework operators, making it difficult for a framework to support many hardware platforms evenly.
    
\end{itemize}

\noindent In a nutshell, the quantization mechanism in deep learning frameworks ensures the accuracy of quantized models but is insufficient to ensure their efficiency on a variety of hardware platforms. The lack of a framework-agnostic toolchain capable of executing pre-quantized models on a variety of hardware platforms limits their deployment at scale.



\subsection{Quantization in Deep Learning Compilers}
\label{sec:intro:dlc}
The emergence of deep learning compilers (hereafter referred to as DL compilers) such as Apache TVM~\cite{tvm}, Facebook Glow~\cite{glow} and Google XLA\cite{xla}, has re-framed the challenge of deploying deep learning models on various hardware platforms. A DL compiler typically converts a model expressed in a framework-specific representation into a common intermediate representation (IR). The DL compiler then successively lowers the graph from the graph-level to the tensor-level. In the graph-level IR, the compiler optimizes the computation graph of the model. In the tensor-level IR, it optimizes the loop structure of tensor operators, which represent the vertices of the computation graph. After successive optimizations, the DL compiler eventually lowers the model to the machine code of a hardware platform using established low-level code generation modules such as LLVM and NVCC. Therefore, compared to deep learning frameworks, deep learning compilers are more effective at handling the multiplicity of front-ends (frameworks) and back-ends (hardware platforms), thereby simplifying the deployment of deep learning models~\cite{tvm,liu2019optimizing}.

However, current work on DL compilers is based mostly on the \texttt{fp32} data type. Although some compilers support the generation and optimization of quantized operators~\cite{cowan2020automatic,glow}, none have focused on compiling and executing pre-quantized models. If a quantized operator is added naively to a deep learning compiler, the new operator must be added across all the IRs. In the graph-level, we need new rules for these new operators. In the tensor-level, we need new computations and possibly new kernel implementations for each platform. The effort could be mitigated somewhat by the overlap between \texttt{fp32} and quantized operator kernel implementation. Nevertheless, the naive approach to adding quantized operators to the compiler would severely limit the pace at which quantization could be applied to deep learning models.

\subsection{Quantized Neural Network Dialect}
This paper proposes \emph{QNN} (Quantized Neural Network) as a graph-level IR dialect that can augment any deep learning compiler. This approach offers an end-to-end solution to read a pre-quantized model and run it across a variety of hardware platforms while reusing most of the existing DL compiler infrastructure. QNN dialect acts as a slightly higher-level IR on top of graph-level IR, specifically designed for handling quantized networks. We add new operators in QNN dialect, but we do not define any graph- or tensor-level optimizations for them. Instead, these operators are lowered to a sequence of DL compiler's existing operators, which already have well-defined graph- and tensor-level optimizations. QNN operators represent quantization constructs at a higher level than the DL compiler's graph-level IR, making QNN a quantization-aware IR.

QNN dialect, therefore, enables reuse of almost all of the existing infrastructure, allowing us to quickly add new hardware platforms, and to focus on kernel implementation of only those operators that are affected significantly by the integer operations (like using VNNI instructions for Intel x86). Additionally, we can define new QNN dialect optimization passes to transform the graph to suit a particular hardware platform, e.g., adding a QNN requantize operator before QNN conv2d to satisfy Intel VNNI \texttt{uint8 x int8} data type requirements. By reusing existing DL compiler infrastructure, QNN dialect reduces the developer efforts needed to efficiently execute pre-quantized models on many hardware platforms. We implemented QNN dialect on top of the open-source deep learning compiler Apache TVM\footnote{https://tvm.apache.org/}.

Specifically, the contributions of this paper are

\begin{itemize}[leftmargin=*]

    \item \boldhdr{QNN Dialect} QNN dialect, a graph-level IR dialect designed to complement deep learning compilers, enables the efficient execution of pre-quantized models on a variety of hardware platforms without cumbersome manual efforts.
    
    \item \boldhdr{Quantization-aware Graph Optimizations} We augment QNN with quantization-aware graph level optimization mechanisms, enabling the graph to meet the requirements imposed by different instruction sets (like $uint8$ x $int8$ for Intel VNNI).
    
    
    \item \boldhdr{Comprehensive Real System Evaluation} We demonstrate that using QNN along with the corresponding graph optimizations, we can compile pre-quantized models from TFLite, MXNet and PyTorch on Intel CPUs, Nvidia GPUs and ARM CPUs equipped on both servers and edge devices and achieve state-of-the-art performance.
    
\end{itemize}

Experiments show that, with the assistance of QNN, a deep learning compiler, specifically Apache TVM, is able to take pre-quantized models defined in TFLite, MXNet and PyTorch, and execute them efficiently on different hardware platforms, with an average speedup of 2.35$\times$, 2.15$\times$ on Intel Xeon Cascade Lake and Nvidia T4 servers, and 1.35$\times$ and 1.40$\times$ on ARM Raspberry Pi3 and Pi4 edge devices, compared to tuned TVM \texttt{fp32} baseline. Generalizability aside, QNN achieves performance comparable to the best state-of-the-art solutions provided by the deep learning frameworks, while also providing better hardware platform coverage than the frameworks. To the best of our knowledge, this is the first unified effort to enable DL compilers for a comprehensive quantized deep learning models support. We have open sourced the QNN work\footnote{https://tvm.apache.org/docs/tutorials/frontend/deploy\_prequantized.html}, which is also used in production.

\section{Problem Setting}
\label{sec:setting}
\subsection{Challenges}
\label{sec:setting:challenges}
Quantization is an essential technique for reducing the compute and memory demand on hardware by replacing the compute data type from \texttt{fp32} to lower-bit integers such as \texttt{int8}. There is an acute need for a software mechanism that can enable developers to take advantage of quantization \emph{easily and rapidly with as little developer effort as possible}. However, building such a mechanism requires that we overcome three major challenges: 

\noindent \boldhdr{Multiple Frameworks} Developers build neural networks in the framework with which they are most comfortable. In practice today, developers in academia and industry use a variety of frameworks to produce pre-quantized models. Therefore, a good solution must be able to handle models from multiple frameworks.

\noindent \boldhdr{Multiple Quantization Approaches} In Section~\ref{sec:intro:back} we briefly discussed various quantization approaches (e.g. symmetric, asymmetric, per-channel, etc.) that developers can choose according to the needs of their application. A good solution must be expressive enough to represent different types of quantization approaches.

\noindent \boldhdr{Multiple Hardware Platforms} Finally, a good solution must be able to run pre-quantized models on a variety of devices across a wide range of compute capabilities. Wherever available, the solution must provide the ability to use  fast integer instructions such as Intel VNNI or Nvidia DP4A instructions.

Currently, developers extend the functionality of deep learning frameworks to process  quantized models. In Section~\ref{sec:intro:framework}, we argued that frameworks have tight coupling of operators and back-end hardware libraries. This leads to operators that are atomic, i.e., they are not decomposable, making it difficult to efficiently execute them on many platforms. We further substantiate this in Table~\ref{tab:frameworks} showing that frameworks have either limited quantization support or limited hardware support. Therefore, existing deep learning frameworks do not present a good solution to tackle all three challenges.

\begin{table}
\begin{tabular}{lllll}
\hline
\multicolumn{1}{|l|}{Quantization Approaches} &
  \multicolumn{1}{l|}{TFLite} &
  \multicolumn{1}{l|}{MXNet} &
  \multicolumn{1}{l|}{PyTorch} &
  \multicolumn{1}{l|}{QNN} \\ \hline
\multicolumn{1}{|l|}{Asymmetric Quantization} &
  \multicolumn{1}{l|}{\cmark} &
  \multicolumn{1}{l|}{} &
  \multicolumn{1}{l|}{\cmark} &
  \multicolumn{1}{l|}{\cmark} \\ \hline
\multicolumn{1}{|l|}{Symmetric Quantization} &
  \multicolumn{1}{l|}{\cmark} &
  \multicolumn{1}{l|}{\cmark} &
  \multicolumn{1}{l|}{\cmark} &
  \multicolumn{1}{l|}{\cmark} \\ \hline
\multicolumn{1}{|l|}{Per-channel Quantization} &
  \multicolumn{1}{l|}{\cmark} &
  \multicolumn{1}{l|}{\cmark} &
  \multicolumn{1}{l|}{\cmark} &
  \multicolumn{1}{l|}{\cmark} \\ \hline
 &
   &
   &
   &
   \\ \hline
\multicolumn{1}{|l|}{Hardware Platforms} &
  \multicolumn{1}{l|}{TFLite} &
  \multicolumn{1}{l|}{MXNet} &
  \multicolumn{1}{l|}{PyTorch} &
  \multicolumn{1}{l|}{QNN} \\ \hline
\multicolumn{1}{|l|}{Nvidia GPU} &
  \multicolumn{1}{l|}{} &
  \multicolumn{1}{l|}{{\cmark}} &
  \multicolumn{1}{l|}{} &
  \multicolumn{1}{l|}{\cmark} \\ \hline
\multicolumn{1}{|l|}{Intel CPU} &
  \multicolumn{1}{l|}{} &
  \multicolumn{1}{l|}{\cmark} &
  \multicolumn{1}{l|}{\cmark} &
  \multicolumn{1}{l|}{\cmark} \\ \hline
\multicolumn{1}{|l|}{ARM CPU} &
  \multicolumn{1}{l|}{\cmark} &
  \multicolumn{1}{l|}{} &
  \multicolumn{1}{l|}{\cmark} &
  \multicolumn{1}{l|}{\cmark} \\ \hline
\end{tabular}
\caption{Frameworks have different degrees of support for uniform $int8$ quantization. More importantly, they do not support all available hardware platforms with good performance. Our solution is designed to eliminate the complexity and effort of supporting all quantization approaches across a variety of hardware platforms with good performance.}
\label{tab:frameworks}
\vspace{-5mm}
\end{table}

\subsection{Observations}
While the state-of-the-art framework-based approaches fail to tackle the challenges of executing pre-quantized models efficiently, the emerging crop of deep learning compilers is promising. DL compilers typically have a framework-agnostic graph-level intermediate representation (IR). We can convert a model from any framework to this IR, solving the first challenge of framework multiplicity. We can also use different types of quantization approaches in a DL compiler, which addresses the second challenge. Finally, DL compilers rely on established code generators like LLVM and NVCC to cover a broad range of hardware platforms, solving the third challenge.

However, an ideal solution must address all these challenges without burdening the developer with extra effort. Because DL compilers were originally designed to compile models in \texttt{fp32} data type, they don't support quantized operators and the corresponding optimizations. As mentioned in Section~\ref{sec:intro:dlc}, simply adding new quantized operators requires a lot of effort across both graph- and tensor-level IRs. This again severely limits how rapidly we can deploy quantized models on to various hardware platforms.

In solving this problem, our key observation is that the computation of a quantized operator can be easily represented as a sequence of simpler operators (in contrast to frameworks where the quantized operators are atomic and cannot be decomposed). We illustrate this with an example of a quantized conv2d operator below, in which $\circ$ denotes a convolution computation.

\begin{equation}
\begin{split}
C_{fp32} & = A_{fp32} \circ B_{fp32} \\
& = [scale_A * (Q_A- zp_A)] \circ [scale_B * (Q_B - zp_B)] \\
& = scale_A* scale_B * Q_C
\end{split}
\label{eq:conv2d}
\end{equation}

\noindent In Equation~\ref{eq:conv2d}, $A_{fp32}$, $B_{fp32}$ are input tensors and $C_{fp32}$ is the conv2d output tensor, all in \texttt{fp32} data type. We first replace the \texttt{fp32} tensors with quantized tensors using Equation~\ref{eq:quant} and then expand the computation. Further, we observe in Equation~\ref{eq:conv2d_terms} that quantized conv2d can be further broken down into four terms. Here, $k$, $c$, $r$ and $s$ represent output channels, input channels, filter height, and filter width respectively, and $n$, $h$, $w$ represent batch size, output height, and output width respectively. We will show the exact sequence of operations in Section \ref{subsec:passes}.

\begin{equation}
\begin{split}
Q_C(n, k, h, w) & = \sum_{c, r, s} Q_A(n, c, h + r, w + s) * Q_B(k, c, r, s)\\
                & - \sum_{c, r, s} zp_A * Q_B(k, c, r, s) \\
                & - \sum_{c, r, s} zp_B * Q_A(n, c, h + r, w + s) \\
                & + \sum_{c, r, s} zp_A * zp_B
\end{split}
\label{eq:conv2d_terms}
\end{equation}

We observe that all the quantized operators can be easily decomposed to simpler, existing operators. In contrast, frameworks have a tight coupling between operators and back-end libraries that prevents such decomposition. Although the decomposition of quantized operators increases the size of the computation graph initially, \emph{we can now reuse the graph- and tensor-level optimizations that the DL compilers provide for models in \texttt{fp32} data type}. For example, we can reuse the graph fusion optimization pass to find and fuse a sequence of operators because we already have the fusion rules for existing operators. Similarly at the tensor-level, we can reuse optimized kernel implementations for many simple operators like integer addition or multiplication, and rely on code generators like LLVM/NVCC, significantly reducing developer effort.

The little effort that developers do spend can be focused on the operators that need attention due to the \texttt{int8} data type. For example, for Intel CPUs, a developer can focus on writing graph-level optimizations to satisfy its $uint8 \times int8$ data type requirement and write kernel implementations using Intel VNNI instructions for quantized conv2d operator. Similarly, for ARMv8, one can focus only on graph- and tensor-level optimizations to use the fast \texttt{int16} multiply-accumulate instructions. The rest of the graph- and tensor-level optimizations can be left to the DL compiler infrastructure.

Based on these observations, our solution is the Quantized Neural Networks (QNN), a graph-level dialect with quantization context that simplifies the efficient execution of pre-quantized models on a variety of hardware platforms as shown in Table~\ref{tab:frameworks}. The QNN dialect enables developers to define a quantized operator simply as a sequence of existing DL compiler operators. Additionally, it allows graph-level optimizations with quantization context to help satisfy the data type requirements imposed by the instruction sets. Therefore, augmented by QNN, DL compilers can make use of quantization across multiple hardware platforms.
\section{Design and Implementation}

\begin{figure}
    \centering
    \includegraphics[width=0.40\textwidth]{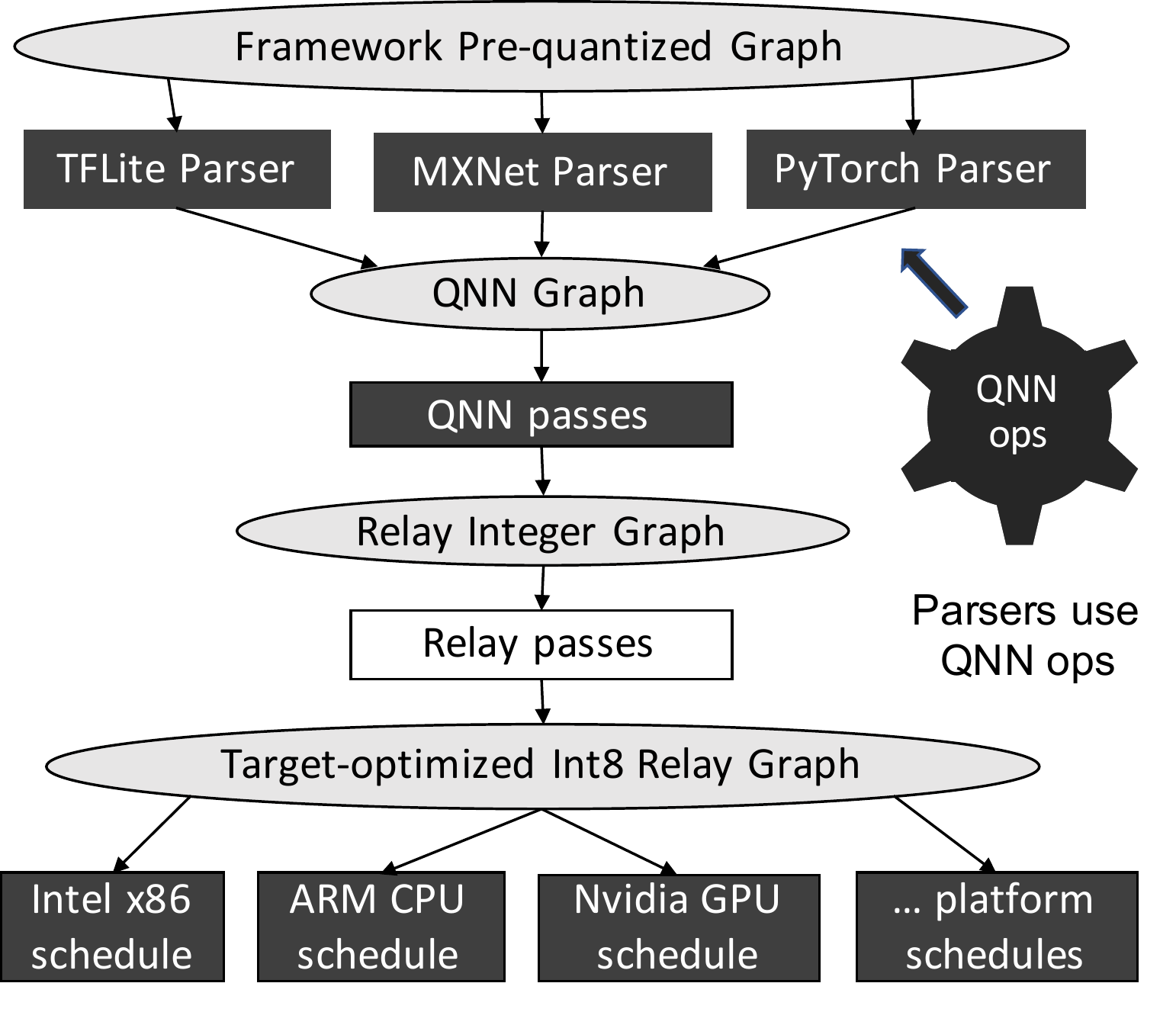}
    \caption{Design of QNN with deep learning compiler - Developer adds new QNN operators with just the description of how they can be lowered to existing graph-level (Relay) operators. Framework parsers use QNN ops to convert the graph to Relay/QNN graph. QNN infrastructure then runs a series of quantization-aware graph-level optimizations, finally produces Relay-only graph (no QNN operators). These operators are then lowered to the machine code using the TVM schedules for different hardware platforms.}
    \label{fig:overview}
    \vspace{-5mm}
\end{figure}

In this section, we first give an overview of QNN dialect and how it fits in the existing DL compiler infrastructure, followed by the design and implementation of different QNN dialect components. We build QNN dialect on top of Apache TVM~\cite{tvm} - an open source deep learning compiler. We will use the TVM terminology to describe the design details with necessary explanation. TVM stack has two levels of IR - a graph-level IR called \emph{Relay}~\cite{roesch2019relay} and a tensor-level IR (hereinafter referred to as tensor IR). QNN dialect is based on the Relay IR, allowing developers to reuse a large portion of existing TVM infrastructure.

An overview of the complete design is shown in Figure~\ref{fig:overview}. A developer first adds a new QNN operator along with the description of how this operator can be lowered to a sequence of existing Relay (or graph-level) operators. Typically, QNN operators correspond to quantized operators defined in the deep learning framework. A framework parser parses the framework model to produce a framework-agnostic graph which is the mix of QNN and Relay operators. Since QNN is a Relay dialect, QNN and Relay operators can co-exist in the same graph. 

This is followed by QNN graph-level optimizations. We implement two QNN passes - QNN Legalize and QNN Canonicalize. QNN optimization passes, just like any graph-level optimization passes, allow graph transformation. However, the key difference is that QNN dialect is a quantization-aware IR, \emph{i.e.} QNN operators have quantization scales, zero points and data type, which is not the case with later Relay passes. The quantization context is helpful to perform hardware-specific transformation (also known as legalization) in QNN Legalize, for example, to satisfy the data type requirements imposed by instruction set. Further, the second QNN pass - QNN Canonicalize pass - converts QNN operators into a sequence of Relay-only operators using a developer-provided sequence. Therefore, QNN Canonicalize pass acts as a boundary after which graph-level quantization context is absent.

From here on, we can reuse the existing TVM infrastructure. We first run Relay optimizations, for example, dead code elimination and graph fusion. After Relay optimizations, each fused operator is then lowered to tensor IR, where it goes through another set of tensor-level optimization passes. Here, a developer can focus on only those operators that require extra attention due to \texttt{int8} data type and customize the kernel implementation for each platform. Finally, the optimized tensor IR is compiled to machine code using off-the-shelf compilers like LLVM/NVCC.

QNN is designed to reduce the developer effort by reusing a large portions of existing DL compiler infrastructure and quickly shifting the developer focus to only those items that are significantly affected by \texttt{int8} data type. Figure~\ref{fig:overview} color codes the new efforts required to augment a DL compiler with QNN using black boxes - QNN operators, framework parser, QNN optimization passes and integer operator schedules.

\subsection{QNN Operators and Framework Parsers}
\label{subsec:ops}

QNN operators act as wrappers, i.e., a developer simply defines how a QNN operator can be represented as a sequence of existing Relay operators. As a result, a developer does not have to add any loop-level tensor IR description of any new QNN operator. This developer provided sequence is used by the QNN Canonicalize pass to convert QNN operators to sequences of Relay-only operators.

In order to narrow down the set of QNN operators to support, we first collected quantized operators from the most widely used frameworks - TFLite, MXNet and PyTorch. We observed that the same operator name can mean different computation manners for different frameworks. For example, TFLite quantized conv2d operator performs \texttt{int8} convolution, requantization of output tensor, ReLU and bias addition in a single operator. The quantized conv2d operator of MXNet goes one step further in aggressive fusion, fusing residual addition operations and folding batch normalization.

\begin{figure}
    \centering
    \includegraphics[width=0.48\textwidth]{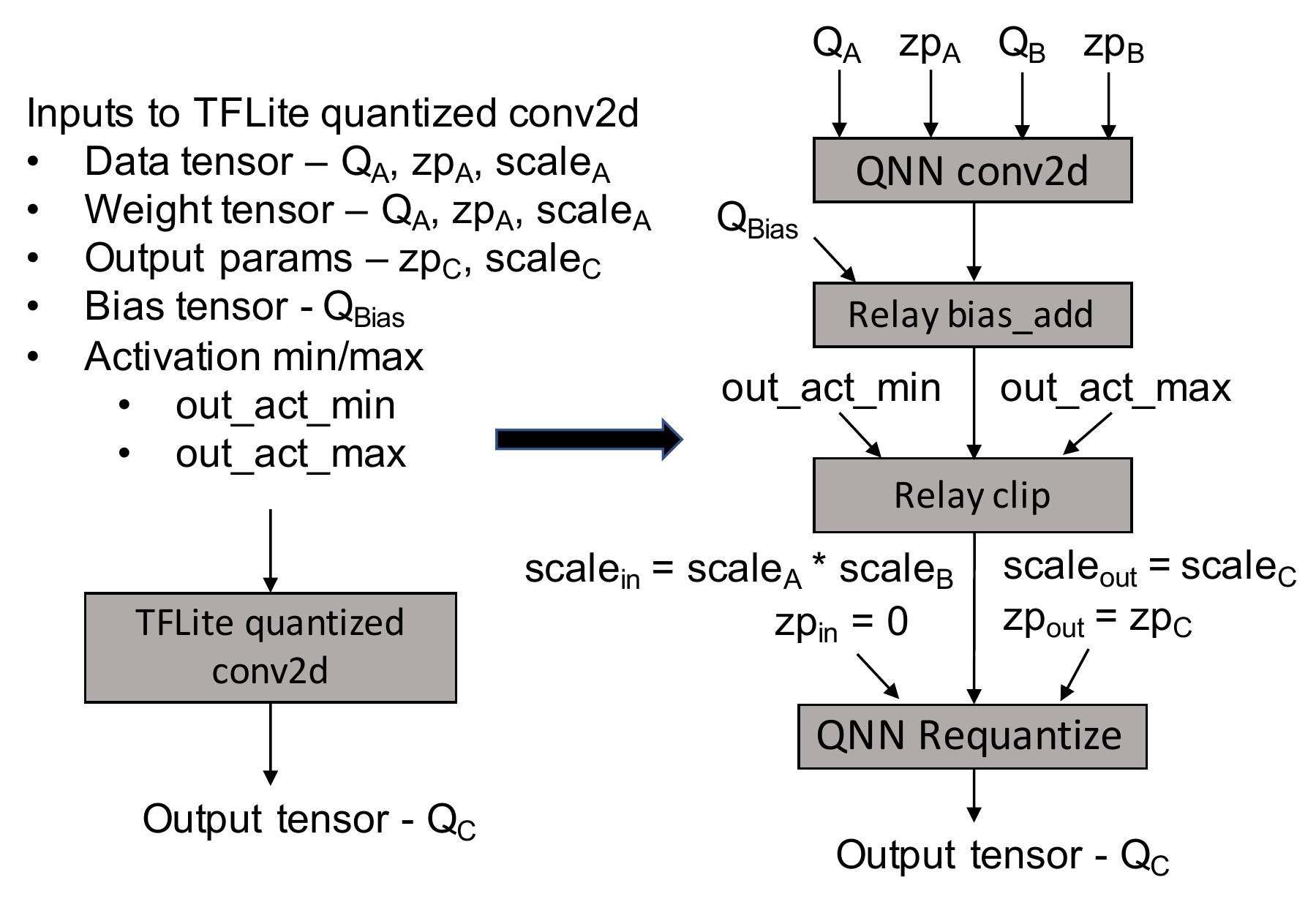}
    \caption{Example of TFLite quantized conv2d operator parsing - TFLite con2d has multiple operators fused internally. We parse it to a sequence of QNN and existing Relay operators - QNN conv2d followed by Relay bias\_add. Then the tensor values are clipped by pre-defined output minimum and maximum values. Finally, we call the QNN requantize operator to go back to int8 datatype.}
    \label{fig:tflite_conv2d}
    \vspace{-4mm}
\end{figure}

To address this computation boundary mismatch, we came up with suitable QNN operators to express different computations for all the mainstream frameworks and all their quantized operators. When necessary, we created finer-grain QNN operators. Essentially, one framework quantized operator can map to a sequence of one or many QNN/Relay operators. We illustrate this idea with the help of an example of conversion of TFLite quantized conv2d operator in Figure~\ref{fig:tflite_conv2d}.

We use a sequence of QNN conv2d, Relay bias\_add, Relay clip and QNN requantize operator to parse the TFLite quantized conv2d operator. We define QNN conv2d computation such that it only handles the quantized tensors and adjustments for zero points as shown in Equation~\ref{eq:conv2d_terms}. For scale handling, we created another QNN operator called requantize, which is extensively used in quantized models to convert one quantized tensor with some scale and zero point to another quantized tensor with another scale and zero point (more details in Section~\ref{subsec:passes}).

A developer, in a similar manner, can follow the computation of a quantized framework operator and can represent it as a sequence of Relay operators with low effort. We added support for different types of quantization approaches in this manner. After the infrastructure was in place, we found that implementing weight per-channel quantization took less than a week of developer effort with QNN dialect requiring no extra work in tensor IR.

\subsection{QNN Canonicalization Pass}
\label{subsec:passes}

Figure~\ref{fig:overview} shows how TVM stack iteratively optimizes the QNN graph after framework parsing is complete. There are two QNN optimization passes - QNN Canonicalize and QNN Legalize. Note that a developer can add more QNN optimization passes if necessary. In this paper, we discuss above two passes which we found sufficient to support a variety of hardware platforms. In this subsection, we discuss QNN Canonicalize. We will discuss QNN Legalize in the next section.

QNN Canonicalize pass converts QNN ops into a sequence of Relay operators using the lowering sequence defined by the developer. Therefore, QNN Canonicalize pass acts as a boundary after which graph-level quantization context is absent, and we reuse existing Relay and tensor-level infrastructure. QNN provides infrastructure where a developer can specify the lowering of a QNN operator to a sequence of Relay operators. This has to be done on an operator-by-operator basis. The difficulty of lowering varies between operators. Here, we show examples of canonicalizing three operators - QNN pooling, QNN conv2d and QNN requantize. The operators are chosen to give a flavor of complexity and share low-level observations and insights about operator designs.

\noindent \textbf{QNN Pooling Operator} - QNN pooling operator canonicalization requires simple lowering. All the framework quantized pooling operators have the same scale and zero point for pooling input and output tensors. This simplifies lowering for quantized pooling as shown below:

\begin{equation}
\begin{split}
B_{fp32} & = AvgPool(A_{fp32}) \\
scale * (Q_B - zp) & = scale * AvgPool(Q_A - zp) \\
Q_B & = AvgPool(Q_A) \\
\end{split}
\end{equation}

\noindent We can skip scale and zero point handling (as they are equal) and just perform average pooling operation. In the pooling operation, we have to be careful about rounding during division and upcast the inputs to \texttt{int16} to avoid overflow/underflow while accumulation. 

\noindent \textbf{QNN Convolution Operator} - As mentioned in Section~\ref{subsec:ops}, we only handle convolution of quantized tensors and adjustments due to zero points (no scale handling) in QNN conv2d operator. As shown earlier in Equation~\ref{eq:conv2d_terms}, QNN conv2d operator can be decomposed into four terms. Each term can be represented using Relay operators. Term 1 is simply Relay conv2d over quantized int8 input tensors. Term 2 can be lowered by performing a reduce sum operation on weight tensor across c, r, s dimension. Term 3 performs a sliding window reduction on input data quantized tensor, which can be represented by pool2d, reduce sum and multiplication operator. And term 4 is just a multiplication of constants. 

\begin{figure}
    \centering
    \includegraphics[width=0.48\textwidth]{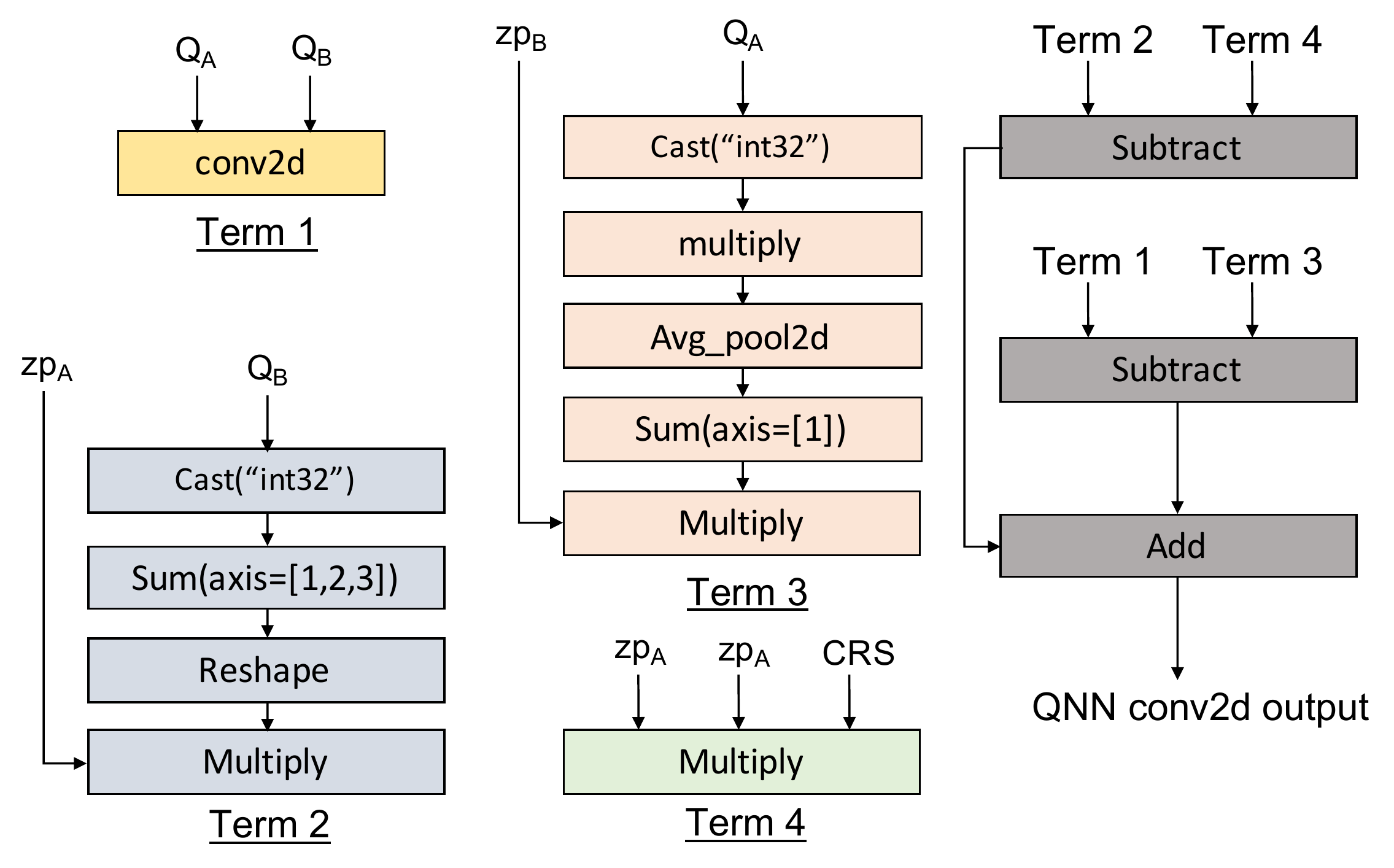}
    \caption{QNN Conv2D canonicalization - QNN Conv2d is lowered across the four terms as described in Equation~\ref{eq:conv2d_terms}. Reshape and cast operators are added to ensure that tensor shapes and data type match while combining the terms. Compile-time constant terms - Term 2 and Term 4 - are subtracted first to perform constant folding.}
    \label{fig:qnn_conv2d}
    \vspace{-5mm}
\end{figure}

We have to be careful about zero points because \texttt{fp32} number $0.0$ is represented by $zero\_point$ in the quantized tensor (which can also be inferred from Equation~\ref{eq:quant}). Therefore, padding a quantized input tensor in QNN conv2d translates to padding the tensor with $zero\_point$. We also have to take care of reshapes to match tensor shapes or allow broadcasting whenever possible. We further observe that term 2 and term 4 are compile-time constants - term 2 is dependent on the weight tensor, which is constant for DNN inference. Specifically, the final QNN-to-Relay canonicalization is shown in Figure~\ref{fig:qnn_conv2d}.

An alternative lowering can be similar to QNN pooling, where we first subtract zero points and then perform conv2d over subtracted tensors as shown in Equation~\ref{eq:conv2d}. However, in this case the tensors have to be upcast to \texttt{int16} before subtracting zero points, causing the final convolution to happen on \texttt{int16} tensors instead of \texttt{int8} tensors. For devices that have fast \texttt{int8} datatype computation support, like Intel VNNI and Nvidia DP4A, this prevents us from using the relevant \texttt{int8} instructions. However for ARMv8 that has a fast \texttt{int16} multiply-accumulate instruction, this lowering gives better performance. QNN Legalize pass, described later, allows this customization for different hardware platforms.

\noindent \textbf{QNN Requantize Operator} - QNN conv2d operator is typically followed by a requantize operator (also shown in Figure~\ref{fig:tflite_conv2d}). Requantize operator changes one quantized tensor representation to another, i.e, we represent the input quantized tensor with new scale and zero point. This can be mathematically written as follows, where $Q_A$ and $Q_B$ are input and output quantized tensors respectively.

\begin{equation}
\begin{split}
scale_B * (Q_B - zp_B) = scale_A * (Q_A - zp_a) \\
Q_B = [(scale_A/scale_B) * (Q_A - zp_a)] + zp_B\\
\end{split}
\label{eq:conv2d_4}
\end{equation}

\noindent Note that scales here are of \texttt{fp32} data type, leading to a floating point multiplication of requantize scale and quantized tensor, which later will have to be converted back to integer data type. This back-and-forth data type conversion can cause severe performance degradation. To solve this problem, we borrow the idea from TFLite requantize operation~\cite{tflite} to use a fixed point multiplication as a proxy for floating point multiplication. An important detail in the implementation is the rounding of fixed point multiplication. Different frameworks choose different rounding methods, which results in minor end-to-end model accuracy differences as shown in the evaluation later.

\subsection{QNN Legalize Pass}

QNN optimization passes, just like Relay passes, allow graph transformations. However, the key difference is that QNN passes have quantization context, \emph{e.g.}, QNN conv2d operators have scale and zero points for the input tensors. The quantization context is helpful to perform hardware-specific graph-IR transformations to satisfy the data type restrictions imposed by the hardware instruction set. We call this pass QNN Legalize pass. Legalization is a common compilation pass that transforms an IR for a specific platform to use the instructions natively supported by the platform. QNN Legalize pass allows developers to easily perform these quantization-aware platform-specific graph optimizations. Its backbone is built on existing Relay infrastructure that allows customization of graph optimization for hardware platforms.

For example, TFLite pre-quantized graphs have $uint8 \times uint8$ inputs for the quantized conv2d operator. However, Intel VNNI instructions-based platforms impose $uint8$ $\times$ $int8$ data type requirement. QNN Legalize pass bridges this gap in a developer-friendly manner by allowing one to insert a requantize operator before the second operand of conv2d, converting the data type from \texttt{uint8} to \texttt{int8}. For ARMv8-based devices, on the other hand, we observe that LLVM performs better code generation if the input tensors are of \texttt{int16} datattype instead of \texttt{int8} datatype, and utilizes fast \texttt{int16} multiply-accumulate instruction ($vmlal$). Therefore, QNN Legalize performs a graph transformation to insert data type upcasting before both operands of the QNN conv2d operator, changing the QNN conv2d input data type to \texttt{int16}.

\subsection{TVM Schedules for Integer Operators}
\label{subsec:schedules}

As shown in Figure~\ref{fig:overview}, after Relay optimization passes have been applied, each fused operator is then lowered to machine code via TVM tensor IR. For many simple operators, like addition or ReLU, that do not have any data reuse, there is not much room for further optimization in addition to relying on off-the-shelf code generators like LLVM/NVCC to get performant machine code. However, operators like conv2d or matmul (matrix multiplication) require specific tensor IR optimizations (also known as a compute and schedule implementation in the context of TVM) to efficiently exploit data reuse. This optimization effort needs to be done for every platform due to drastic architectural differences.

QNN infrastructure quickly shifts the developer focus to only those operators that need extra attention due to integer computations. For example, we can reuse existing TVM schedules for integer pooling, vector addition, ReLU etc. This is in contrast to frameworks, where a new quantized operator (which handles scale and zero points internally) need to be implemented separately. For the operators significantly affected by the integer computation, we can write specific schedules for them to utilize the fast integer instructions provide by the hardware to get desirable performance. We present our observations regarding the usage of these instructions across both server and edge devices.

\noindent \emph{\textbf{Intel VNNI}} - TVM relies on off-the-shelf code generators to generate good quality code. However in some cases, it might be difficult for LLVM to use the right instructions automatically. For example, Intel VNNI instruction performs a vector dot-product of 4 \texttt{int8} values and accumulate them into \texttt{int32}, potentially achieve 4$\times$ speedup over \texttt{fp32} computation. However, by now LLVM is still unable to detect this macro pattern from LLVM IR to replace with proper Intel VNNI instructions. Therefore, in this case a developer can directly embed the LLVM intrinsics in the TVM tensor IR. We used this feature to write high performance TVM schedules for integer convolution operators for Intel CPUs.

\noindent \emph{\textbf{ARM Edge Devices}} - In contrast to Intel VNNI, the Raspberry Pi edge devices, based on ARMv8 architecture, do not have hardware support for fast \texttt{int8} dot product instruction. However, ARMv8 ISA has a fast \texttt{int16} multiply-accumulate instruction ($vmlal$) that can perform dot product of 2 16-bit values and accumulate in 32-bit. We observe that LLVM picks up the $vmlal$ instructions for code generation if the input tensors are of \texttt{int16} datatype instead of \texttt{int8}. Therefore, we use QNN Legalize pass to insert the up-casting operations before the QNN conv2d operators for ARMv8-based devices.

\noindent \emph{\textbf{Nvidia GPUs}} - Similar to Intel VNNI, Nvidia has a DP4A instruction to speedup 8-bit integer computation. Recently, Nvidia has also introduced tensor cores to achieve even further speedup. In this work, we leverage the already existing Nvidia DP4A TVM schedule. Note that given TVM abstractions, in future, a developer can just focus on the TVM schedule for convolution using Tensor Cores, and easily replace the DP4A schedule with the new schedule. Writing TVM schedule using Tensor Core is beyond the scope of this paper. 

Overall, QNN is designed to augment DL compilers, in our case, Apache TVM, to deploy pre-quantized models efficiently across many hardware devices with low developer effort. In cases where we need extra attention due to specific integer instructions, QNN can still reduce a significant portion of developer's time and effort by reusing the existing TVM infrastructure.

\section{Evaluation}
This section evaluates our proposed QNN solution by answering the following questions:
\begin{enumerate}[leftmargin=*]
    \item As a sanity check, is QNN able to compile pre-quantized models to achieve similar model accuracy numbers compared to the framework solutions?
    \item What is the performance of QNN-compiled pre-quantized models, in comparison to the original models in \texttt{fp32}?
    \item Can QNN get the on-par, if not better, performance on pre-quantized models compared to the framework solutions while covering more hardware platforms than frameworks?
    \item How does QNN perform in compiling a newly designed pre-quantized model?
\end{enumerate}
\subsection{Experimental Setup}
\label{sec:methodology}

\noindent \boldhdr{Frameworks} We evaluate QNN across all the available pre-quantized models in TFLite (version 1.13)~\cite{tflite_models}, MXNet (version 1.6)~\cite{mxnet_models} and PyTorch (version 1.4)~\cite{pytorch_models}. We implement QNN on top of open-source Apache TVM (version 0.6)~\cite{tvm}. Note that different frameworks support different sets of pre-quantized models as listed in Table~\ref{tab:frameworks}, while QNN-augmented TVM is able to compile all of them.

\noindent \boldhdr{Server Platforms} We evaluate QNN on two server platforms on Amazon EC2 - Intel 24-core Xeon Cascade Lake CPU equipped at EC2 C5.12xlarge instance and Nvidia T4 GPU at EC2 G4.xlarge instance. Both processors have hardware support for speeding up $int8$ computations - Intel VNNI and Nvidia DP4A instructions. The Nvidia T4 GPU also has recently introduced Tensor Cores. However, for this evaluation we only use DP4A instructions. Using Tensor Cores is an orthogonal effort to this paper (refer to Section~\ref{subsec:schedules}).

\noindent \boldhdr{Edge Devices} We evaluate QNN on two popular edge devices - Raspberry Pi3 (in-order ARM Cortex A53) and Raspberry Pi4 (out-of-order ARM Cortex A72). In contrast to our server platforms, Raspberry Pi devices do not have any fast $int8$ computation instructions. Instead, they have 16-bit multiply-accumulate instructions ($vmlal$) that leads to better data packing in registers.

\subsection{Deploying QNN across Frameworks}

First of all, we evaluate the effectiveness of QNN in achieving a wide framework coverage. Since each framework has its own preferred choice of quantization approach (asymmetric, symmetric, per-channel), we simultaneously evaluate the ability of QNN to represent different quantization approaches. Specifically, we compare the accuracy of pre-quantized models achieved by the frameworks and QNN-augmented TVM stack. We measure the accuracy over 10k images from the Imagenet validation dataset~\cite{imagenet} and show the findings in Table \ref{tab:mxnet}.

\begin{table}
\begin{tabular}{|l|l|l|l|l|}
\hline
\multirow{2}{*}{Quantized Model} & \multicolumn{2}{l|}{Top1 Accuracy (\%)} & \multicolumn{2}{l|}{Top5 Accuracy (\%)} \\ \cline{2-5} 
  & Baseline & QNN-TVM & Baseline & QNN-TVM \\ \hline
\multicolumn{5}{c}{MXNet Pre-quantized Models}\\ \hline
resnet-18 & 69.76	& 69.86	& 89.02	& 89.05    \\ \hline
resnet-50 & 76.13	& 76.16	& 92.6	& 92.73    \\ \hline
resnet-50-v1b & 76.66	& 76.56	& 92.6	& 92.6    \\ \hline
resnet-101 & 77.13	& 76.97	& 93.06	& 93.09    \\ \hline
resnet-152 & 75.99	& 75.75	& 92.52	& 92.12    \\ \hline
inception-v3 & 77.84	& 77.28	& 93.52	& 93.32    \\ \hline
inception-bn & 71.96	& 71.79	& 90.38	& 90.25    \\ \hline
mobilenet-v1 & 71.27	& 71.13	& 90.09	& 90.16    \\ \hline
mobilenet-v2 & 70.35	& 70.14	& 89.45	& 89.52    \\ \hline
\multicolumn{5}{c}{TFLite Pre-quantized Models}\\ \hline
inception-v1 & 70.1 & 69.6 & 89.8 & 89.5    \\ \hline
inception-v2 & 73.5 & 73.3 & 91.4 & 91.3    \\ \hline
inception-v3 & 77.5 & 77.3 & 93.7 & 93.6    \\ \hline
inception-v4 & 79.5 & 79.6 & 93.90 & 94.2    \\ \hline
mobilenet-v1 & 70.0 & 70.1 & 89.0 & 89.0    \\ \hline
mobilenet-v2 & 70.8 & 70.9 & 89.9 & 90.1    \\ \hline
\multicolumn{5}{c}{PyTorch Pre-quantized Models}\\ \hline
resnet-18 & 69.49 & 69.63 & 88.67 & 88.47 \\ \hline
resnet-50 & 75.88 & 75.84 & 92.64 & 92.67 \\ \hline
inception-v3 & 77.65 & 77.28 & 93.36 & 93.18 \\ \hline
googlenet & 69.59 & 69.37 & 89.34 & 89.28 \\ \hline
mobilenet-v2 & 70.43 & 70.61 & 89.48 & 89.44 \\ \hline
\end{tabular}
\caption{QNN achieves accuracy parity across all the mainstream frameworks -- MXNet, TFLite and PyTorch -- pre-quantized models.}
\label{tab:mxnet}

\end{table}

We observe that QNN achieves accuracy parity for all pre-quantized models across the frameworks with minor differences. As explained in Section~\ref{subsec:passes}, these differences mainly attribute to the rounding operations in fixed point multiplication of the requantize operator. Different frameworks use different rounding methods, leading to small differences in final accuracy.

\subsection{Deploying QNN Across Server and Edge Devices}

QNN is designed to enable efficient deployment of pre-quantized models across a variety of hardware platforms with different types of computing capabilities. In this subsection, we evaluate QNN effectiveness in performance speedup and memory footprint reduction, when it is deployed across our server and edge devices.

\begin{figure}
    \centering
    \includegraphics[width=0.48\textwidth]{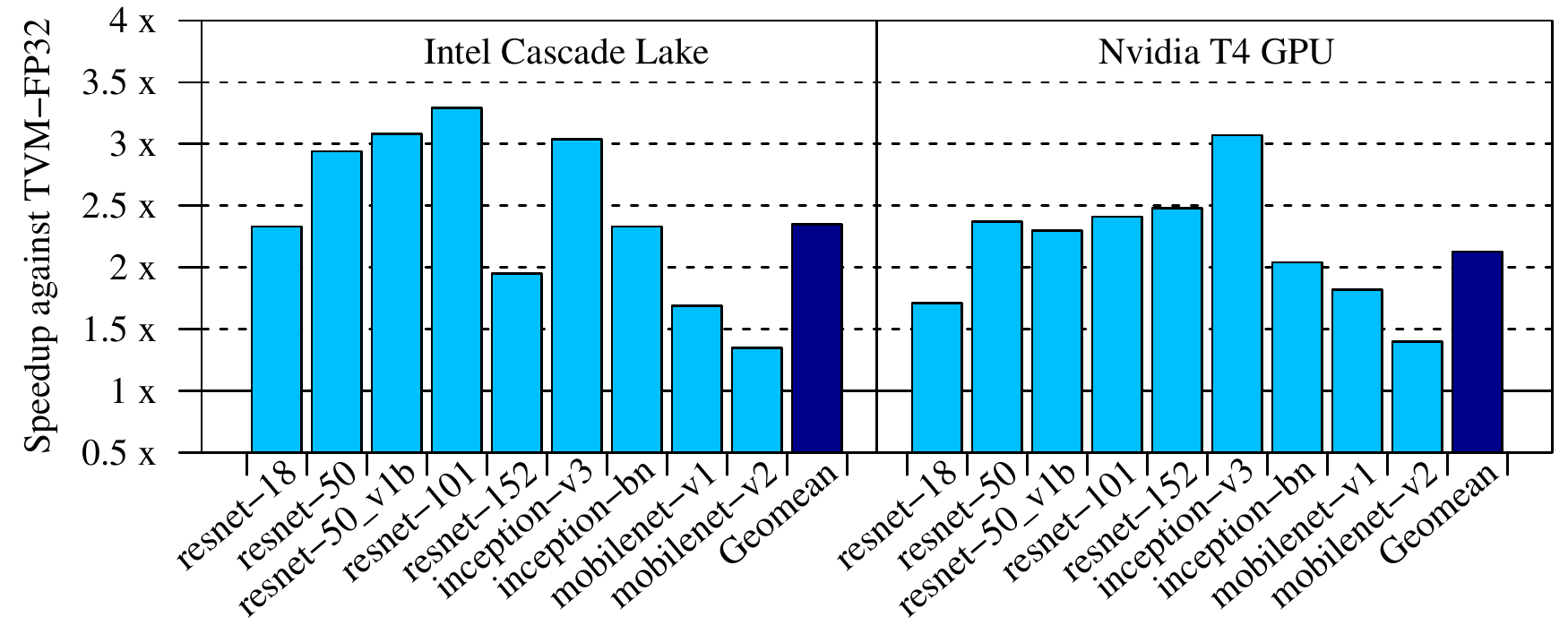}
    \caption{QNN on Server - QNN takes advantage of fast \texttt{int8} instructions to achieve significant speedup against TVM \texttt{fp32} tuned baseline for both Intel Cascade lake and Nvidia T4 servers.}
    \label{fig:money_server}
\end{figure}

\begin{figure}
    \centering
    \includegraphics[width=0.48\textwidth]{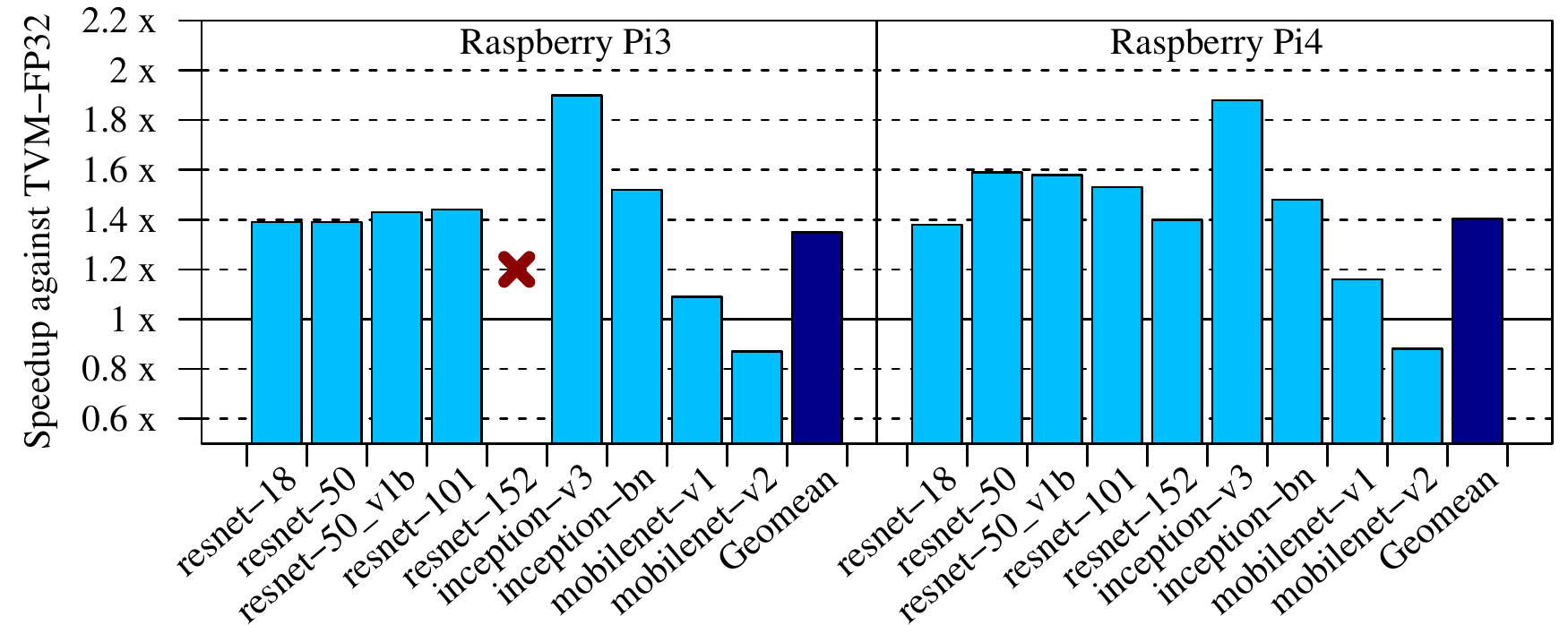}
    \caption{QNN at edge - QNN enables easy deployment across the edge devices as well. We see performance speedup for both in-order (A53) and out-of-order (A72) ARM cores. The red cross shows that \texttt{fp32} resnet-152 model execution was out of memory, while QNN \texttt{int8} execution succeeded.}
    \label{fig:money_edge}
    \vspace{-5mm}
\end{figure}

\begin{figure*}
    \centering
    \includegraphics[width=\textwidth]{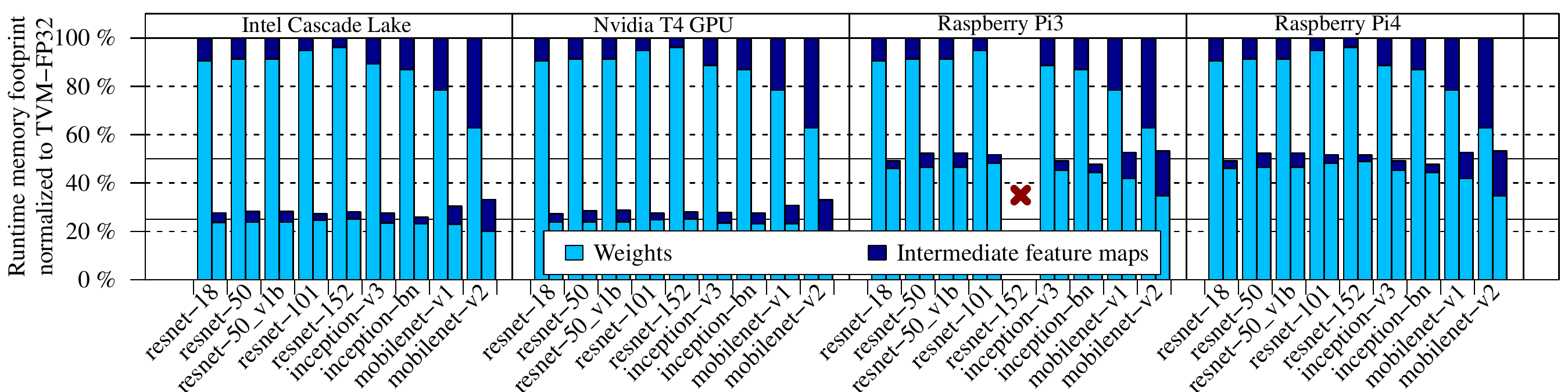}
    \caption{Breakdown of runtime memory footprint across weights and intermediate feature maps. Intermediate feature also play a considerable role in deciding the total memory footprint reduction. For Intel and Nvidia servers, QNN-\texttt{int8} reduces the total memory footprint by 67-74\%. For ARM, we currently upcast weights to \texttt{int16} to use ARM fast $vmlal$ instruction, resulting in 50\% (or 2$\times$) footprint reduction.}
    \label{fig:memory}
    \vspace{-5mm}
\end{figure*}

\noindent \boldhdr{Performance Improvements} In this experiment, we compile the original MXNet models (in \texttt{fp32}) and their counterpart pre-quantized model (in \texttt{int8}) using QNN-augmented TVM stack. We execute each compiled model for 2000 images (batch size 1) and measure the average end-to-end latency. We also perform auto-tuning~\cite{autotvm} to ensure high performance for both original and pre-quantized models.

We compare the performance of TVM-compiled original models (referred to as TVM-\texttt{fp32}) and QNN-augmented TVM-compiled MXNet pre-quantized models (referred to as QNN-\texttt{int8}). There are two reasons for choosing this baseline. First, no framework can run across all the platforms we are targeting, making TVM-\texttt{fp32} a good baseline. Second, MXNet has the largest number of available pre-quantized models amongst all the frameworks, enabling wider model coverage evaluation. The same observation applies to pre-quantized models of TFLite and PyTorch. We present the results for server-class platforms and edge-devices in Figure \ref{fig:money_server} and Figure \ref{fig:money_edge} respectively. We observe very low standard deviation amongst 2000 runs, and therefore omit the error bars in the barplots.

For servers, we observe significant performance improvements for both Intel Cascade Lake CPU and Nvidia T4 GPU servers as shown in Figure \ref{fig:money_server}. Both processors have support for fast \texttt{int8} dot-product operations - Intel VNNI and Nvidia DP4A instructions. We also observe lower than expected speedup for resnet-152 and mobilenet models. For resnet-152, MXNet decided to keep the batch normalization operator in \texttt{fp32} to retain accuracy, causing frequent data type conversions and hurting performance. For mobilenet models, TVM stack currently lacks good depthwise convolution schedules (kernel implementation) using fast \texttt{int8} instructions. Overall, we observe that QNN-\texttt{int8} achieves an average speedup of 2.35$\times$ and 2.13$\times$ for Intel Cascade Lake CPU and Nvidia T4 GPU respectively compared to TVM-\texttt{fp32}.

Similarly for edge devices, we observe significant speedups as shown in Figure~\ref{fig:money_edge}. In contrast to servers, our edge devices do not have fast \texttt{int8} instructions, leading to lower speedups than observed in servers. However, these devices have fast $int16$ multiply-accumulate instructions ($vmlal$). We observe that LLVM generates highly vectorized and vector register-packed code using $vmlal$ instructions, efficiently utilizing convolution data reuse and achieving better performance than \texttt{fp32} models. Additionally, we observe that TVM-\texttt{fp32} resnet-152 model goes out of memory due to large weight size in Pi3 (shown by cross in the figure). QNN-\texttt{int8} on the other hand, due to smaller memory footprint, can execute the model. Similar to servers, mobilenet models observe sub-optimal performance due to the lack of good TVM schedules for depthwise convolution operator. Overall, QNN-\texttt{int8} achieves an average speedup of 1.35$\times$ and 1.40$\times$ for ARM Raspberry Pi3 and Pi4 respectively compared to TVM-\texttt{fp32}.

Note that, although mobilenet models show lower than expected performance speedup, QNN allows developers to just focus on implementing better TVM schedules for depthwise convolution operator, while reusing the existing TVM infrastructure for both graph- and tensor-level optimizations, enabling rapid deployment with less developer efforts.

\noindent \boldhdr{Memory Footprint Reduction} In this experiment, we evaluate QNN effectiveness in reducing the runtime memory footprint. We compare the total runtime memory footprint for TVM-\texttt{fp32} and QNN-\texttt{int8} models for all the hosted MXNet models across all hardware platforms. We present the findings in Figure~\ref{fig:memory}, showing QNN-\texttt{int8} total memory footprint as a percentage of total TVM-\texttt{fp32} footprint. We also break down total memory footprint into 2 categories - weights (also known as parameters) and intermediate feature maps (also known as activations or intermediate outputs). In contrast to prior works that only show weight memory footprint reduction~\cite{tflite, quant1}, we analyze total memory footprint.

We observe that different models have different memory footprint reduction depending on the contribution of intermediate feature maps to total memory footprint. As opposed to weights that have been pre-quantized to \texttt{int8} and achieve close to 4$\times$ memory footprint reduction, the intermediate features maps that can also be in \texttt{int32} data type, observe less than 4$\times$ memory reduction. For example, mobilenet models have larger contribution of intermediate feature maps, overall reducing the footprint to 33\% (or 3$\times$ footprint reduction). 

We also observe that in edge devices, weights of QNN-\texttt{int8} see only 50\% memory footprint reduction (much less than expected 75\% or 4$\times$). This is because we upcast the weights to \texttt{int16} to use ARM fast \texttt{int16} multiply-accumulate $vmlal$ instruction. Note that this overhead is unavoidable. TVM stack runs a constant evaluation pass that converts the \texttt{int8} pre-quantized weights to \texttt{int16} at compile-time. If we disable constant evaluation optimization pass and keep the weights in \texttt{int8}, the tensor with upcast weights in \texttt{int16} datatype will still be the part of total memory footprint, still keeping the total memory footprint same. Therefore, we measure total memory footprint to accurately assess the total memory footprint reduction.


\subsection{QNN Comparison with Frameworks}

Next, we compare the performance between QNN and frameworks of executing pre-quantized models. As shown in Table~\ref{tab:frameworks}, frameworks do not have efficient pre-quantized model execution support for all hardware platforms. Therefore, we execute all the hosted pre-quantized models for each framework on the hardware platforms it supports, and compare the performance with the same models compiled and executed by QNN-augmented TVM. Therefore, this evaluation compares our work against the best available baseline. We do not show comparison on GPU platforms, because TFLite and PyTorch do not support pre-quantized model execution on GPUs, while MXNet supports but with suboptimal performance~\cite{mxnet_jun}.

\noindent \boldhdr{MXNet Framework} MXNet framework presents the best baseline for Intel CPUs because it relies on Intel DNNL that has hand-written x86 assembly implementations. For example, Intel DNNL uses Intel VNNI instructions to achieve high performance for \texttt{int8} data type convolution or matrix multiplication operators. The performance of executing quantized models on Nvidia GPUs in MXNet framework is under-optimized, and MXNet does not have a backend to generate high performance ARM machine code. Therefore, we can not compare QNN with MXNet on Nvidia and ARM platforms but just focus on Intel CPUs.

\begin{figure}
    \centering
    \includegraphics[width=0.48\textwidth]{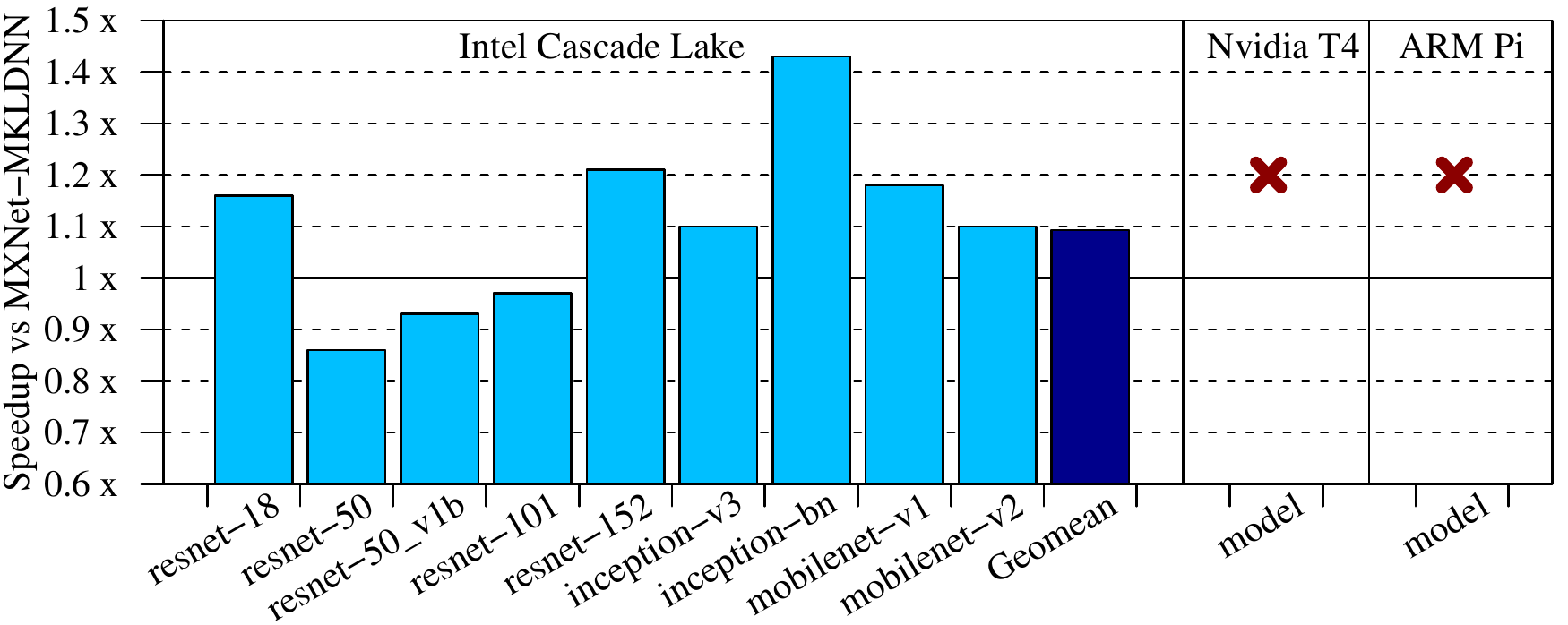}
    \caption{MXNet vs QNN - QNN achieves an average speedup of 1.09$\times$ against highly hand-tuned Intel DNNL execution of pre-quantized models. The red cross signifies that MXNet does not have good support of running pre-quantized models on ARM and Nvidia devices.}
    \label{fig:best_baseline_mxnet}
\end{figure}

Figure~\ref{fig:best_baseline_mxnet} summarizes the end-to-end performance comparison between MXNet and our solution. Overall, we observe QNN-\texttt{int8} achieves 1.09$\times$ speedup against MXNet on Intel Cascade Lake CPU, with a maximum of 1.43$\times$ for inception-bn. We observe slowdown for resnet-50, and resnet-50\_v1b models. We suspect this is because Intel DNNL has customized the optimization for resnet models due to their popularity. However, for other less popular models, we observe significant speedup.

\begin{figure}
    \centering
    \includegraphics[width=0.48\textwidth]{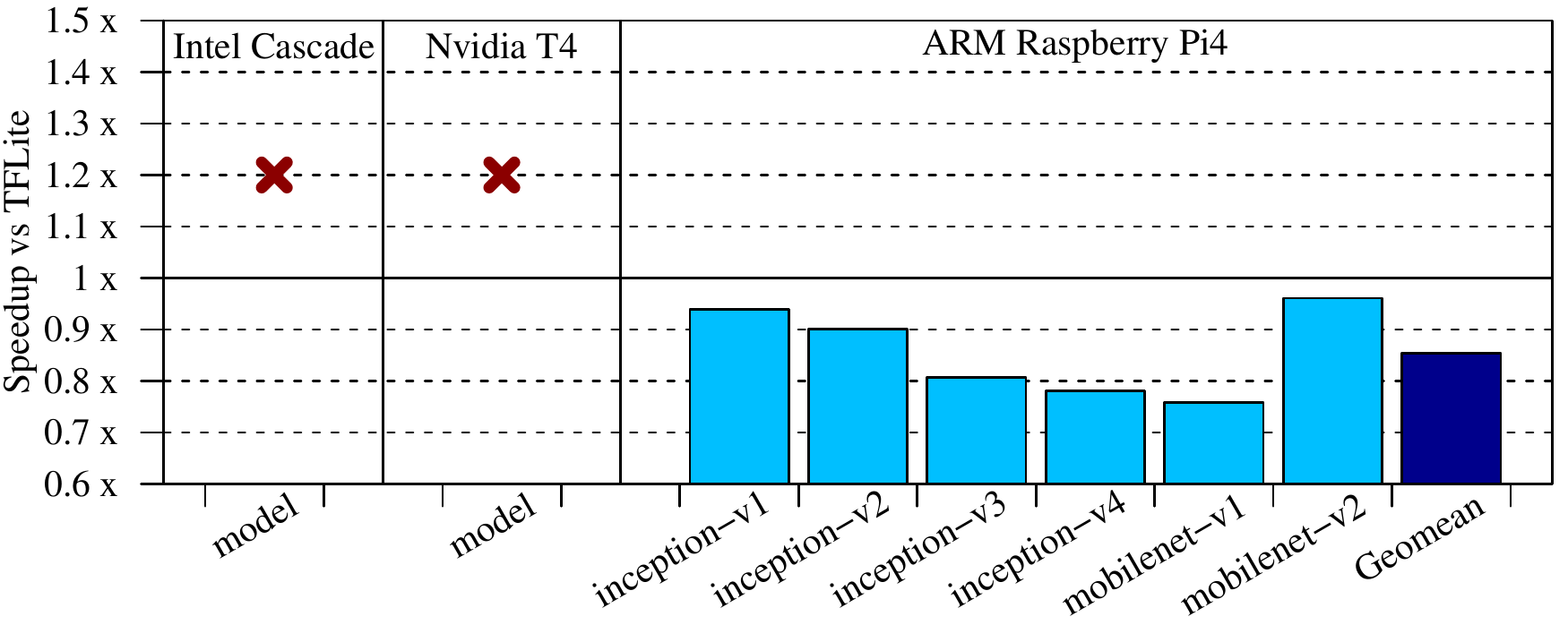}
    \caption{TFLite vs QNN - Currently QNN is on average 15\% slower than TFLite for pre-quantized models on ARM devices because of hand-written tuned assembly implementations for operators. The red cross signifies that TFLite does not have good support of running pre-quantized models on Intel and Nvidia devices.}
    \label{fig:best_baseline_tflite}
    \vspace{-6mm}
\end{figure}

\noindent \boldhdr{TFLite Framework} TFLite framework has been designed to mainly focus on edge devices. TFLite uses hand-tuned implementation for \texttt{int8} operator on ARM devices. However, given its scope, TFLite does not have good performance on Intel CPUs and Nvidia GPUs. For example, we observe that our solution is over 10$\times$ faster on Intel Cascade Lake CPUs against TFLite execution for resnet-50 model. Therefore, we do not show comparison of QNN-augmented TVM with TFLite on Intel and Nvidia devices. In this experiment, we measure the performance of QNN-augmented TVM and TFLite execution for all TFLite pre-quantized models on Raspberry Pi4.

The findings of this experiment are shown in Figure~\ref{fig:best_baseline_tflite}, demonstrating as QNN-augmented TVM speedup against TFLite performance. We observe that TFLite is faster than QNN-augmented TVM for all cases. Overall, QNN-augmented TVM is 14\% (or 1.16$\times$) slower than TFLite on average. This is because of hand-tuned assembly implementations for \texttt{int8} conv2d and depthwise operators in TFLite. With the help of improving \texttt{int8} ARM schedules which is out of scope of this paper, we should be able to close this gap.

\begin{figure}
    \centering
    \includegraphics[width=0.48\textwidth]{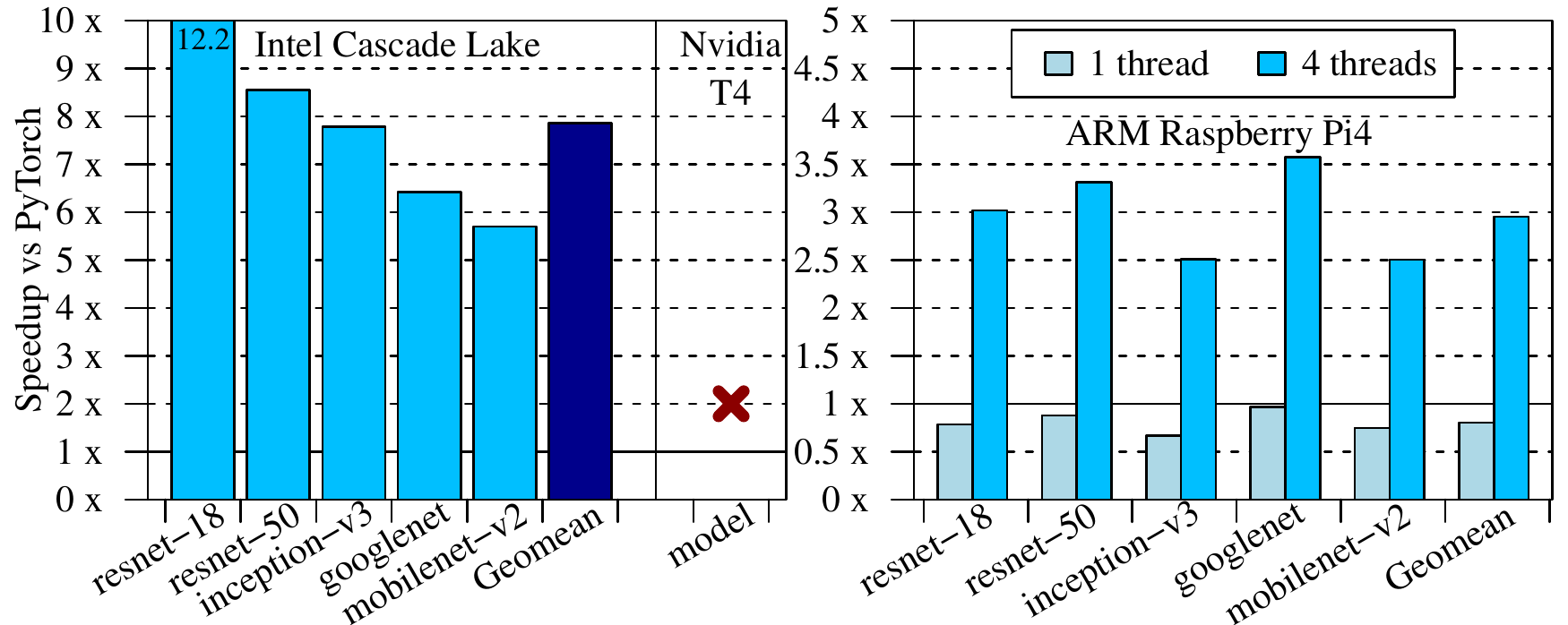}
    \caption{PyTorch vs QNN - QNN achieves $7.85\times$ speedups on Intel Cascade Lake against PyTorch-FBGEMM. PyTorch-QNNPACK does not support multi-threading on Raspberry Pi4. QNN is 20\% slower on average for single thread and 2.95$\times$ faster for four threads execution.}
    \label{fig:best_baseline_pytorch}
    \vspace{-6mm}
\end{figure}

\noindent \boldhdr{PyTorch Framework} PyTorch relies on hand-written assembly implementations - FBGEMM~\cite{fbgemm} for Intel CPUs and on QNNPACK~\cite{qnnpack} for ARM mobile devices - to achieve high performance for pre-quantized models. PyTorch does not have any support for executing pre-quantized models efficiently on Nvidia GPUs. In this experiment, we measure end-to-end performance of PyTorch and QNN-augmented TVM for all hosted PyTorch pre-quantized models on Intel Cascade Lake CPU and ARM Raspberry Pi4. The findings of this experiment are shown in Figure~\ref{fig:best_baseline_pytorch}, demonstrating speedup of QNN-augmented TVM normalized to PyTorch performance.

We observe that QNN-augmented TVM achieves high speedups across all the models against PyTorch-FBGEMM execution on Intel Cascade lake servers, with up to 12.2$\times$ for resnet-18 and an average of 7.85$\times$. PyTorch community could work on improving these numbers, but in general, it highlights the effectiveness of QNN-augmented TVM in quickly achieving high performance.

We found that PyTorch does not use multiple threads for QNNPACK on non-Android or iOS devices including Raspberry Pi4. Therefore, we show comparison with TVM single thread performance in addition to TVM four threads performance. We observe that, similar to our findings with TFLite, there is room of improvement for better ARM schedules overall. Overall, QNN is 20\% slower in the single-thread apples-to-apples comparison but 2.95$\times$ faster if all four cores are used.


\subsection{QNN Effectiveness on a New Model}
Lastly, we present a scenario where we use QNN to compile and execute a new unseen pre-quantized model. Our new model is an in-house built keyword detection model, which is executed on a resource-constrained edge device, to detect specific keywords or triggers in human speech. Typically, these models are executed very frequently on the device to capture the keywords. Therefore, it is imperative to have both low latency and resource utilization for this model so that we can co-execute other applications on the same device simultaneously.

To reduce the latency, we employed framework quantization to quantize the model and then compiled the model for ARM Cortex A53 (Raspberry Pi3) edge device using QNN-augmented TVM (QNN-\texttt{int8}). For evaluation, our baseline is the \texttt{fp32} model compiled and executed via TVM (TVM-\texttt{fp32}). We observe that, without any extra developer efforts, QNN-\texttt{int8} was 2.0$\times$ and 2.7$\times$ faster than TVM-\texttt{fp32} for single-threaded and multi-threaded execution (4 threads), while achieving 50\% total runtime memory footprint reduction and no accuracy loss. This shows that a QNN-augmented deep learning compiler is effective in rapidly deploying new models on new devices with low developer efforts.

\section{Related Work}

\noindent \boldhdr{8-bit Integer Quantization} 8-bit integer quantization is the most widely adopted quantization technique because of prevalence of \texttt{int8} data type computation support in the commodity hardware platforms. 8-bit integer quantization has also been shown to retain model accuracy with relatively less efforts compared to more aggressive quantization. TFLite 8-bit integer quantization, designed primarily for ARM-based edge devices, was one of the first large-scale effort and set the industry standard for integer quantization~\cite{Jacob_2018_CVPR}. There is a large body of prior work to retain model accuracy for quantization, which can be broken into two categories - Post-training quantization and Quantization-aware training~\cite{mxnet-mkl, nvidia-talk, tflite}. Post-training quantization starts from an already trained \texttt{fp32} model and uses calibration on a cross-validation dataset to find suitable quantization parameters~\cite{mxnet-mkl,nvidia-talk}. Research has shown that post-training quantization achieves good accuracy for 8-bit quantization. More aggressive quantization requires quantization-aware training. Our approach can leverage all of these research to get a pre-quantized model, and eases its deployment on many hardware platforms.

\noindent \boldhdr{Deep Learning Frameworks} Deep learning frameworks, like Tensorflow, PyTorch and MXNet, have been at the forefront of deep learning revolution. These frameworks rely on pre-built backend hardware libraries for high-performance machine code - Intel DNNL~\cite{dnnl}, Nvidia CuDNN~\cite{cudnn} and ARM ACL~\cite{acl}. These backend libraries have hand-optimized kernel implementations to achieve high performance on different platforms. Most of these frameworks use \texttt{int8} quantization to support lightweight model inference on commodity hardware. A user also has a choice of using symmetric, asymmetric or per-channel quantization approaches to find a suitable performance-accuracy tradeoff. However, as shown in Table~\ref{tab:frameworks}, frameworks are not well suited to deploy the models across many hardware platforms. QNN dialect is designed to leverage the extensive work done by the frameworks in quantizing the models and retaining accuracy, and solve the challenges to ease the deployment of pre-quantized models on many platforms with low developer efforts.

\noindent \boldhdr{Deep Learning Compilers} Deep learning compilers have gained popularity in past few years to design a flexible approach for optimizing and deploying deep learning models. Some of the DL compilers support optimization and code generation for multiple hardware platforms - Apache TVM~\cite{tvm}, Glow~\cite{glow} and XLA~\cite{xla}. On the other hand, there are compilers focusing on only one class of hardware platforms - Intel nGraph~\cite{ngraph}, ARM NN~\cite{armnn} and Nvidia TensorRT~\cite{nvidia, cudnn}. There has also been recent efforts in improving the compiler design for supporting deep learning hardware accelerators, for example, in Apache TVM~\cite{tvm}, Glow~\cite{glow}, PlaidML~\cite{plaidml} with stripe IR~\cite{stripe} and MLIR~\cite{mlir}. Quantization support in deep learning compilers has picked up pace very recently and therefore good support is limited to the compilers focusing on a single class of hardware platform - nGraph, TensorRT and ARM NN. This limitation prevents us from rapidly deploying models on many platforms with single unified toolchain. Our work on the QNN dialect is designed to complement the existing DL compilers to solve this challenge. We observe that QNN-augmented DL compiler compiles pre-quantized models from multiple frameworks, supports different types of quantization approaches and enables deployment on many hardware platforms with low developer efforts.

\section{Conclusion}
In this paper, we offer a universal solution to the challenge of efficiently executing pre-quantized models from a variety of frameworks across a variety of hardware platforms while keeping developer effort low. We note that most deep learning frameworks help developers quantize models but fall short at supporting the efficient execution of models on multiple platforms. This is due to the tight coupling of framework operators and back-end hardware libraries, which hinders the use of quantization. We tackle the problem using the notion of a deep learning compiler enhanced with a new graph-level dialect called Quantized Neural Network (QNN). The QNN dialect provides a quantization context that can augment any existing deep learning compiler (\emph{e.g.} Apache TVM, Glow, XLA). QNN simplifies the effort of efficiently executing pre-quantized models on variety of hardware devices. We observe that our QNN-augmented deep learning compiler achieves speedups of 2.35$\times$, 2.15$\times$, 1.35$\times$ and 1.40$\times$ on Intel Xeon Cascade Lake CPUs, Nvidia Tesla T4 GPUs, ARM Cortex-A CPUs on Raspberry Pi3 and Pi4 respectively relative to \texttt{fp32} execution. QNN also achieves comparable performance against the state-of-the-art framework's solutions for executing pre-quantized models while providing much better coverage of hardware platforms.

\bibliographystyle{IEEEtran}
\bibliography{references}

\end{document}